\documentclass[twocolumn,a4paper,eqsecnum,english,a4paper,showpacs,preprintnumbers,amsmath,amssymb,prb]{revtex4}
\usepackage[latin1]{inputenc}
\usepackage{float}
\usepackage{graphicx}
\usepackage{amssymb}

\makeatletter

\usepackage{float}
\makeatletter
\usepackage{float}
\makeatletter
\usepackage{dcolumn}
\usepackage{bm}

\makeatother

\usepackage{babel}
\makeatother
\begin{document}

\title{Resonant inelastic scattering spectra at 
the Ir $L$-edge in Na$_2$IrO$_3$}

\author{Jun-ichi Igarashi$^{1}$ and and Tatsuya Nagao$^{2}$}

\affiliation{
 $^{1}$Faculty of Science, Ibaraki University, Mito, Ibaraki 310-8512,
Japan\\
$^{2}$Faculty of Engineering, Gunma University, Kiryu, Gunma 376-8515,
Japan
}

\date{\today}

\begin{abstract}
We analyze resonant x-ray scattering (RIXS) spectra in 
Na$_2$IrO$_3$ on the basis of
the itinerant electron picture. Employing a multi-orbital 
tight-binding model on a honeycomb lattice, we find that the 
zigzag magnetic order is the most stable with sizable energy gap 
in the one-electron 
band within the Hartree-Fock approximation. We derive the RIXS 
spectra, which are connected to the generalized 
density-density correlation function. 
We calculate the spectra as a function of excitation energy $\omega$,
within the random phase approximation. The spectra
consist of the peaks with $\omega$ $<20$ meV, and of the peaks 
with $0.4<\omega<0.8$ eV.
The former peaks are composed of four bound states in the 
density-density correlation function, and may be identified 
as the magnetic excitations,
while the latter peaks are composed of sixteen bound states below 
the energy continuum of individual electron-hole pair excitations, 
and may be identified as the excitonic excitations. 
The calculated spectra agree qualitatively with the recent RIXS experiment.
\end{abstract}


\maketitle
\section{Introduction\label{sect.1}}
Resonant inelastic x-ray scattering (RIXS) has attracted much
interest as a useful tool probing the elementary excitations 
in the materials\cite{Ament2011-rmp,Dean2015,Fatale2015,Tohyama2015}.
It could directly access to the $d$ states in transition-metal compounds  
by using the $L$-edge resonance. 
The magnetic excitations have been clearly 
detected due to recent instrumental improvement of the energy resolution
\cite{Braicovich2009,Braicovich2010,Guarise2010}.
In this respect, RIXS could be compared with inelastic neutron
scattering (INS).
In particular, RIXS provides a valuable tool for the study of magnetic
order when some isotopes show strong neutron absorption.
Iridium, for instance, is the case where its main isotopes are
strong neutron absorbers.
 
Recently, the elementary excitations in iridium oxides, 
such as CaIrO$_3$\cite{Sala2014}, Sr$_2$IrO$_4$\cite{J.Kim2012}, 
and Na$_2$IrO$_3$\cite{Gretarsson2013-1,Gretarsson2013-2}, 
have been investigated by RIXS experiments.
Such $5d$ transition-metal compounds have drawn much attention, since
their physical properties would be quite different from those 
of the $3d$ transition-metal compounds because of
the competition between the large spin-orbit interaction (SOI) and the 
Coulomb interaction. 
Among them, we focus on Na$_2$IrO$_3$ in this paper. 
Its crystal structure belongs to the space group $C2/m$ 
\cite{Choi2012,Ye2012,Lovesey2012},
where Ir$^{4+}$ ions constitute approximately two-dimensional
honeycomb lattice with a Na ion located at its center.
It is revealed to be a magnetic insulator with a zigzag spin order
on the honeycomb lattice
\cite{Choi2012,Ye2012,Singh2010,Comin2012,Liu2011}.
In the RIXS experiment,
the spectra as a function of excitation energy $\omega$ show 
peaks with $\omega < 35$ meV\cite{Gretarsson2013-1,Gretarsson2013-2}, 
which may be associated with the
magnetic excitations. The INS experiment has detected similar peaks
much lower region below 6 meV\cite{Choi2012}.
In addition, there exist peaks with $0.4<\omega<0.8$ eV 
\cite{Gretarsson2013-1,Gretarsson2013-2}, which may be associated with
the excitonic excitations.

The spin order as well as the magnetic excitations have been 
studied theoretically on the basis of the localized electron picture. 
In these studies, spin models of Kitaev-Heisenberg 
type with the spin-orbital coupled isospin $j_{\rm eff}=\frac{1}{2}$
\cite{Kitaev2006,Chaloupka2010,Kimchi2011,Chaloupka2013,Sizyuk2014}
and of more generic types of models\cite{Katukuri2014,Rau2014,Yamaji2014}
have been introduced.
On the other hand, the band structure calculations based on the density
functional theory (DFT) have been carried out 
\cite{Mazin2012,Foyevtsova2013,Kim2014}. 
By making the self-interaction correction (SIC) to the DFT,
it is found that the zigzag spin order is realized as the ground state, 
and that the energy band has an energy gap $\sim 0.5$ eV \cite{Kim2014}. 
Although such itinerant-electron approaches look promising,
the elementary excitations have not been calculated yet along this line.
In our previous paper, employing a tight-binding model on a honeycomb
lattice, we have calculated the generalized density-density correlation
function within the random phase approximation (RPA) on a zigzag spin 
ordered ground state given by the Hartree-Fock approximation (HFA)
\cite{Igarashi2016}. 
Though the calculated correlation function has captured some of the
qualitative aspects of the low energy excitation scheme
exhibited in the RIXS data, a direct comparison between 
the experiment and the theoretical RIXS spectra may be a
next requirement.

As regards the RIXS spectra, no theoretical attempt has been done
within the itinerant electron picture. Therefore, in this paper,
we formulate the RIXS spectra connected to the density-density
correlation function on the basis of the itinerant model
introduced in our previous work \cite{Igarashi2016}.
With the revised parameter values of the tight-binding model, 
we calculate the RIXS spectra within the HFA-RPA scheme.
Note that the HFA-RPA scheme has worked well 
when the antiferromagnetic long-range order is established, 
even if the system is strongly correlated, such as La$_2$CuO$_4$
\cite{Nomura2005,Nomura2015} and Sr$_2$IrO$_4$ \cite{Igarashi2014-3}. 
In the present case, the HFA to the 
tight-binding model leads to a magnetic insulator with the 
zigzag spin order in the ground state, although the energy differences 
from those of the N\'{e}el or stripy orders are obtained as small as 
$\sim 50$ meV per Ir ion. We have the energy gap $\sim 0.8$ 
eV in the one-electron energy band with the conduction band 
composed mainly of the $J_{\rm eff}=\frac{1}{2}$ states.
These results are consistent with the SIC-DFT calculation \cite{Kim2014}.

In formulating the RIXS spectra, we adopt the fast 
collision approximation(FCA), which is  
justified when the Ir $2p$-level life-time
broadening width is larger than the relevant excitation energy.
Then, the RIXS spectra are 
connected to the density-density correlation function. 
We evaluate the formula of the RIXS 
spectra as a function of $\omega$ 
for momentum transfer ${\bf q}$ along symmetry directions within the RPA.
We obtain the peaks for $\omega<20$ meV, 
which are originated from four bound states 
in the density-density correlation function.
We also obtain the peaks for $0.4<\omega<0.8$ eV, which are 
originated from sixteen bound states in  
the density-density correlation function
below the energy continuum of individual 
electron-hole pair excitations.
The calculated spectra are in qualitative agreement with the 
RIXS experiments\cite{Gretarsson2013-1,Gretarsson2013-2}.

The present paper is organized as follows. In Sec. 2,
we introduce a multi-orbital tight-binding model and carry out the HFA 
to the model.
In Sec. 3, introducing the dipole transition,
we formulate the RIXS spectra in terms of the density-density correlation
function within the FCA, which is expressed within the RPA.
In Sec. 4, we present the numerical calculations.
Section 5 is devoted to the concluding remarks. 

\section{Model and the Hartree-Fock approximation\label{sect.2}}
\subsection{Description of model}
Neglecting the crystal distortion, we assume
each Ir ion in Na$_2$IrO$_3$ resides around the center of oxygen octahedra.
The energy level of the $e_g$ orbitals of Ir atom is about 2-3 eV higher 
than that of the $t_{2g}$ orbitals due to the crystal electric field 
of IrO$_6$. Therefore, taking account of only $t_{2g}$ orbitals,
we employ a multi-orbital tight-binding model on a honeycomb lattice,
which is defined by
\begin{equation}
 H = H_{\rm SO}+H_{\rm I}+ H_{\rm kin},
\end{equation}
with
\begin{eqnarray}
H_{\rm SO} & = & \zeta_{\rm SO}\sum_{i}\sum_{nn'\sigma\sigma'}
 d_{in\sigma}^{\dagger}({\bf L})_{nn'}
 \cdot({\bf S})_{\sigma\sigma'}d_{in'\sigma'}, \\
H_{\rm I} & = & 
  U\sum_{i,n} n_{in\uparrow}n_{in\downarrow} \nonumber \\
  &+&
 \sum_{i,n<n'\sigma}[U' n_{in\sigma}n_{in'-\sigma}
                 + (U'-J) n_{in\sigma}n_{in'\sigma}] \nonumber\\
 &+&J\sum_{i,n\neq n'} (d_{in\uparrow}^{\dagger}d_{in'\downarrow}^{\dagger}
                     d_{in\downarrow}d_{in'\uparrow}
                    +d_{in\uparrow}^{\dagger}d_{in\downarrow}^{\dagger}
                     d_{in'\downarrow}d_{in'\uparrow}), \nonumber\\
\label{eq.H_Coulomb} \\
H_{\rm kin} & = & \sum_{\left\langle i,i'\right\rangle }
\sum_{n,n'\sigma}[\hat{T}_{i,i'}]_{n,n'}d_{in\sigma}^{\dagger}d_{i'n'\sigma}
+ {\rm H.c.},
\end{eqnarray}
where the annihilation ($d_{in\sigma}$) and creation 
($d_{in\sigma}^{\dagger}$) operators are for the $5d$ electron with orbital 
$n$ ($=yz,zx,xy$) and spin $\sigma$ at the Ir site $i$, and 
$n_{in\sigma}\equiv d_{in\sigma}^{\dagger}d_{in\sigma}$.

The $H_{\rm SO}$ stands for the SOI of $5d$ electrons with 
${\bf L}$ and ${\bf S}$ denoting the orbital and spin angular momentum 
operators, respectively. 
The $H_{\rm I}$ represents the Coulomb interaction
among the $t_{2g}$ electrons, satisfying $U=U'+2J$ \cite{Kanamori1963}.
We use the values $\zeta_{\rm SO}=0.45$ eV, $U=1.4$ eV, and $J/U=0.15$ 
in the following calculation. 
Similar values of the parameters have been utilized for Ir atom in
Sr$_2$IrO$_4$ \cite{Igarashi2014-3}.

\begin{figure}
\includegraphics[width=8.0cm]{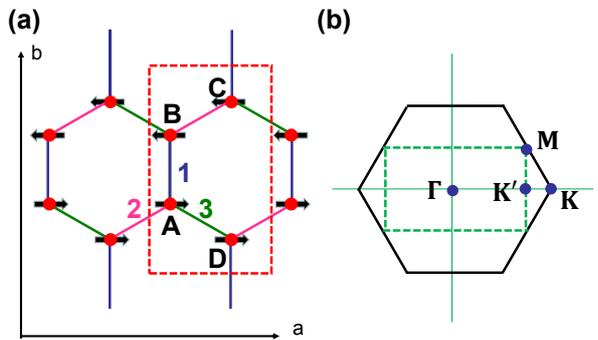}
\caption{\label{fig.unit}
(a) Unit cell for the zigzag ordered phase in a honeycomb lattice.
It consists of Ir atoms A, B, C, and D, which is enclosed by a broken line.
The attached numbers 1, 2, 3 indicate the types of bonds.
Arrows illustrate the zigzag spin ordering pattern.
(b) Corresponding magnetic Brillouin zone, which is enclosed by a 
broken line.}
\end{figure}

The $H_{\rm kin}$ stands for the kinetic energy of $5d$ electrons with
the hopping matrix $\hat{T}_{i,i'}$ between the sites $i$ and $i'$.
The summation $\langle i,i'\rangle$ is restricted within the
nearest neighbor Ir ions. 
There exist two kinds of transfer mechanisms between 
the Ir $5d$ orbitals. The first one is 
a direct transfer between the Ir $5d$ orbitals,
which may be described by means of the Slater-Koster parameters, 
$V_{dd\sigma}$, $V_{dd\pi}$ ($=-2V_{dd\sigma}/3$),
and $V_{dd\delta}$ ($=V_{dd\sigma}/6$)\cite{Slater1954}. 
We may evaluate $V_{dd\sigma}=-0.511$ eV by applying Harrison's 
procedure\cite{Harrison}. 
The second one is an indirect transfer via oxygen $2p$-orbitals.
Since the energy required for transferring the electron 
from oxygen $2p$ levels to Ir $5d$ levels
is rather large, the effective transfer between $5d$ orbitals may be
estimated by the second-order perturbation. Hence the hopping matrix 
$\hat{T}_{i,i'}^{(pd)}$ 
may be expressed in a matrix form with the bases $n=yz$, $zx$, $xy$ in order:
\begin{equation}
 \hat{T}_{i,i'}^{(pd)} 
= \left(\begin{array}{ccc}
                        0 & 0 & t_p \\
                        0 & 0 &  0 \\
                       t_p & 0 &  0 
                          \end{array} \right), 
                  \left(\begin{array}{ccc}
                        0 & 0 &  0 \\
                        0 & 0 &  -t_p \\
                        0 & -t_p &  0 
                          \end{array} \right), 
                  \left(\begin{array}{ccc}
                        0 & -t_p &  0 \\
                        -t_p & 0 &  0 \\
                        0 & 0 &  0 
                          \end{array} \right), 
\end{equation} 
where $\langle i,i'\rangle$ belongs to bonds 1, 2, and 3, respectively,
with the bonds defined in figure \ref{fig.unit} (a).  
The $t_p$ may be as large as $\sim 0.25$ eV, 
which value is similar to the one used in the study of Sr$_2$IrO$_4$. 
The opposite sign had been assigned
to $t_p$ in our previous paper \cite{Igarashi2016}.

\subsection{Hartree-Fock Approximation\label{sect.2.2}}
We carry out the HFA by following the procedure given in 
\cite{Igarashi2016}. 
In that process, we rewrite the Hamiltonian in terms of the Fourier 
transformed operator with the wave vector ${\bf k}$ within the first 
magnetic Brillouin zone (MBZ) 
shown in figure \ref{fig.unit}(b): 
\begin{equation}
 d_{\lambda n\sigma}({\bf k}) 
= (4/N)^{\frac{1}{2}} 
\sum_{i \in \lambda}d_{in\sigma}
       \textrm{e}^{-i{\bf k}\cdot{\bf r}_i} .
\label{eq.Fourier}
\end{equation}
Here, $\lambda$ specifies one of the four sublattices A, B, C, and D,
and $N$ is the number of Ir ions. 
The first MBZ is divided into 
$40\times 40$ meshes in the numerical calculation.
The parameter values are set $\zeta_{\rm SO}=0.45$ eV, $U=1.4$ eV, 
$J/U=0.15$, $V_{dds}=-0.511$ eV, and $t_p=0.25$ eV.

The staggered magnetic moment is known to be directing to the $a$ axis
by the experiment \cite{Liu2011}.
Fixing the staggered moment parallel to the $a$ axis, we carry out the
HFA to evaluate the energies of the zigzag, N\'{e}el, and stripy 
ordered states.
We find the zigzag ordered state is the most stable.
Its energy per Ir ion is 0.09 eV and 0.04 eV smaller than 
that of the N\'{e}el ordered state 
and that of the stripy ordered state, respectively. 
The orbital and spin moments align 
anti-parallel to each other at each site, and their magnitudes are
$|\langle L_a \rangle|=0.51$ and $|\langle S_a \rangle|=0.16$, 
respectively, where $\langle X \rangle$ represents the ground
state average of operator $X$.
Hence the total magnetic moment is evaluated as 0.19 $\mu_{B}$,
which is comparable to the experimental value of 0.22 $\mu_{B}$.

\begin{figure}
\includegraphics[width=8.0cm]{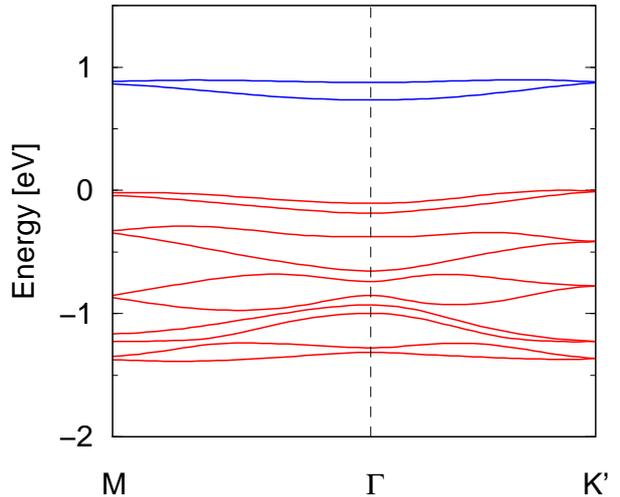}
\caption{\label{fig.one-electron}
One-electron energy as a function of ${\bf k}$ along symmetry directions
within the HFA. Parameters are set $V_{dds}=-0.511$ eV, $t_p=0.25$ eV, 
$\zeta_{\rm SO}=0.45$ eV, $U=1.4$ eV, and $J/U=0.15$.
The origin of energy is set at the top of the valence band. 
Each band is doubly degenerate. }
\end{figure}

Figure \ref{fig.one-electron} shows the one-electron energy 
as a function of ${\bf k}$ along symmetry directions in the 
zigzag ordered state. Each line is doubly degenerate. 
Hence there exist four states per unit cell in the conduction band.
The energy gap is found as large as 
$\sim 0.8$ eV, and the dependence on ${\bf k}$ is rather weak. 
The "$j_{\rm eff}=\frac{1}{2}$" states have the largest weight for the conduction band. 
The band gap seems overestimated in comparison with the SIC-DFT calculation\cite{Kim2014}.

\section{Formulation of RIXS spectra\label{sect.3}}
\subsection{Dipole transition at the Ir L edge}
In the dipole transition, the $2p$ core-electron is excited
to the $5d$ states by absorbing photon at the $L$ edge 
(and the reverse process). This process may be described by the interaction 
\begin{eqnarray}
H_{x}&=&
\sum_{i}\sum_{n,\sigma,j_{\rm{c}}, m,\alpha}
w(n\sigma;j_{\rm{c}}m;\alpha) \nonumber\\
&\times&\sum_{\bf{q}}
[d_{i n\sigma}^{\dagger}p_{i j_{\rm{c}}m}
c_{\alpha}({\bf q}) \textrm{e}^{i{\bf q}\cdot{\bf r}_i} 
+{\rm H.c.}],
\label{eq.dipole}
\end{eqnarray}
where $c_{\alpha}({\bf q})$ is the annihilation operator of photon 
with momentum ${\bf q}$ and polarization $\alpha$. 
The $p_{i j_{\rm{c}}m}$ is the annihilation operator of 
core electron with the angular momentum $j_{\rm{c}}$ 
($=\frac{3}{2}$, $\frac{1}{2}$),
and magnetic quantum number $m$ at site $i$. 
The $w(n\sigma;j_{\rm{c}}m;\alpha)$ represents the matrix 
elements of the $2p\to 5d$ transition, 
which explicit values for $\alpha=x, y,$ and $z$ are listed 
in Table I of \cite{Igarashi2014-3}.
In the Fourier transform representation, 
(\ref{eq.dipole}) may be rewritten as
\begin{eqnarray}
H_{x}&=&\sum_{\lambda,n,\sigma,j_{\rm{c}}, m,\alpha}
w(n\sigma;j_{\rm{c}}m;\alpha) \label{eq.Hx1}\\
&\times&\sum_{\bf{k},\bf{q}}
[d_{\lambda n\sigma}^{\dagger}([{\bf k+q}])
p_{\lambda j_{\rm{c}}m}({\bf k})
c_{\alpha}({\bf q}) \textrm{e}^{-i{\bf G}\cdot{\bf a}_{\lambda}} 
+{\rm H.c.}],
\nonumber
\end{eqnarray}
with
\begin{equation}
 p_{\lambda j_{\rm c} m}({\bf k}) 
=  (4/N)^{\frac{1}{2}}
\sum_{i}p_{i j_{\rm c} m}
  \textrm{e}^{-i{\bf k}\cdot{\bf r}_i} ,
\end{equation}
where ${\bf k}$ is defined on the first MBZ in the honeycomb lattice.
The photon momentum ${\bf q}$ is now regarded as the photon momentum
projected onto the two-dimensional $ab$ plane. 
Here $[{\bf k+q}]$ indicates the vector reduced back to the MBZ by 
a reciprocal vector ${\bf G}$, that is,
${\bf k+q}=[{\bf k+q}]+{\bf G}$, and ${\bf a}_{\lambda}$ is defined 
by $(0,0)$, $(0,1)$, $(\sqrt{3}/2,3/2)$, and
$(\sqrt{3}/2,-1/2)$ in units of $a$ (the nearest neighbor distance)
for $\lambda=$ A, B, C, and D, respectively.

\subsection{Second-order optical process and the scattering geometry} 
The RIXS spectral intensity may be expressed by the second-order 
optical process,
\begin{eqnarray}
 W(\omega_{\rm i},q;\alpha_{\rm i},\alpha_{\rm f}) 
&=& 2\pi\sum_{f'}
 \left | \sum_{n}
 \frac{\langle\Phi_{f'}|H_x|\Phi_n\rangle
       \langle\Phi_n|H_x|\Phi_{\rm i}\rangle}
 {\omega_{\rm i}+\epsilon_{\rm g}
 -\epsilon_n+i\Gamma_{\rm c}}\right|^2 \nonumber \\
 &\times& \delta(\omega_{\rm i}+\epsilon_{\rm g}
-\omega_{\rm f}-\epsilon_{f'}) ,
\label{eq.second}
\end{eqnarray}
where the initial state $|\Phi_{\rm i}\rangle$ may be expressed by
$c_{\alpha_{\rm i}}^{\dagger}({\bf q}_{\rm i})|0\rangle|{\rm g}\rangle$ 
with $|{\rm g}\rangle$ denoting the ground state of the matter 
with energy $\epsilon_{\rm g}$, and $|0\rangle$ denoting the 
vacuum state with photon,
respectively. The intermediate state  $|\Phi_n\rangle$ may be expressed 
by $|0\rangle|n\rangle$ with $|n\rangle$ denoting the intermediate state 
of the matter with energy $\epsilon_n$, and $\Gamma_{\rm c}$ stands 
for the life-time broadening width of the $2p$-core hole state. 
The state $|\Phi_{f'}\rangle$ may be expressed by
$c_{\alpha_{\rm f}}^{\dagger}({\bf q}_{\rm f})|0\rangle|f'\rangle$ with 
$|f'\rangle$ denoting the excited state of the matter with energy 
$\epsilon_{f'}$. 
The incident photon has momentum and energy $q_{\rm i}
=({\bf q}_{\rm i},\omega_{\rm i})$ 
with polarization $\alpha_{\rm i}$, while the scattered photon 
has momentum and energy $q_{\rm f}=({\bf q}_{\rm f},\omega_{\rm f})$ 
with polarization $\alpha_{\rm f}$. 
The momentum and energy transferred to the matter are accordingly given
by $q=q_{\rm i}-q_{\rm f}=({\bf q},\omega)$.

\begin{figure}
\includegraphics[width=8.0cm]{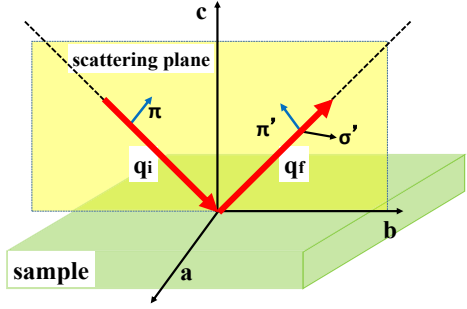}
\caption{\label{fig.geometry}
Horizontal scattering geometry. The scattering plane is assumed to be 
perpendicular to the $ab$ plane, and is including the $b$ axis.}
\end{figure}

Most of RIXS experiments have been carried out in a horizontal scattering 
geometry with the scattering angle $2\theta$ set to be close 
to 90$^{\circ}$.
In the following analysis, we tentatively assume that the scattering plane 
is perpendicular to 
the plane of honeycomb, and is including the $b$ axis, as illustrated
in figure \ref{fig.geometry}.  Note that only a few degrees of 
tilt of the scattering plane could sweep the entire Brillouin zone, 
since $\omega\sim 11.2$ keV at the Ir $L_3$ edge. 
In the coordinate frame with $a$, $b$, and $c$ axes, 
the polarization vectors are given by
\begin{eqnarray}
 \mbox{\boldmath{$\alpha$}}_{\rm i} &=& 
\begin{array}{lcl}
\left(0,\frac{1}{\sqrt{2}},\frac{1}{\sqrt{2}}\right), 
&\mbox{for} & \pi \ \mbox{polarization}, \\
\end{array} \label{eq.pol.i} \\
 \mbox{\boldmath{$\alpha$}}_{\rm f} &=& \left\{
\begin{array}{lcl}
(1,0,0), 
&\mbox{for} & \sigma' \ \mbox{polarization}, \\
\left(0,-\frac{1}{\sqrt{2}},\frac{1}{\sqrt{2}}\right),
&\mbox{for} & \pi' \ \mbox{polarization}.\\
\end{array} \right. \label{eq.pol.f}
\end{eqnarray}
Note that the $t_{2g}$ states, $yz$, $zx$, and $xy$, as well as 
the polarizations in $w(n\sigma;j_{\rm c} m;\alpha)$ are defined 
in the cubic coordinate frame.

\subsection{Fast collision approximation}
Since the value of $\Gamma_{\rm c}$ ($\sim 2.5$ eV) is much larger than 
a variation of $\epsilon_n$ with $n$,
the energy denominator of (\ref{eq.second}) could be factored out
in a reasonable accuracy. 
This procedure is called as the FCA, 
and (\ref{eq.second}) may be rewritten as
\begin{eqnarray}
&& \hspace*{-1.0cm}
 W(\omega_{\rm i},q;\mbox{\boldmath{$\alpha$}}_{\rm i},
                    \mbox{\boldmath{$\alpha$}}_{\rm f}) 
= 2\pi
  |R(\omega_{\rm i},E_0)|^2\sum_{f'} \nonumber\\
&\times& 
 \left|\sum_n
    \langle \Phi_{f'}|H_x|\Phi_n\rangle
    \langle\Phi_n|H_x|\Phi_{\rm i}\rangle\right|^2 
       \delta(\omega+\epsilon_{\rm g}-\epsilon_{f'})
\label{eq.Wfast}
\end{eqnarray}
with
\begin{equation}
 R(\omega_{\rm i},E_0)
=\frac{1}{\omega_{\rm i}-E_0+\epsilon_{\rm 2p}(j_{\rm c})
                         +i\Gamma_{\rm c}}, \label{eq.R}
\end{equation}
where $E_0$ stands for a typical energy of the conduction band.
Using the Fourier transform representation, (\ref{eq.Wfast}) 
is rewritten as, 
\begin{eqnarray} 
&& \hspace*{-1.0cm}
W(\omega_{\rm i},q;\mbox{\boldmath{$\alpha$}}_{\rm i},
                   \mbox{\boldmath{$\alpha$}}_{\rm f})
 \nonumber\\
&&\hspace*{-1.0cm}=
 |R(\omega_{\rm i},E_0)|^2
 \hat{M}^{\dagger}(\mbox{\boldmath{$\alpha$}}_{\rm i},
                   \mbox{\boldmath{$\alpha$}}_{\rm f};j_{\rm c})
\hat{Y}^{+-}(q)
 \hat{M}(\mbox{\boldmath{$\alpha$}}_{\rm i},
         \mbox{\boldmath{$\alpha$}}_{\rm f};j_{\rm c}),
\label{eq.second2}
\end{eqnarray} 
where
\begin{eqnarray}
 \left[\hat{M}(\mbox{\boldmath{$\alpha$}}_{\rm i},
               \mbox{\boldmath{$\alpha$}}_{\rm f};j_{\rm c})
 \right]_{\xi\xi'}
&=& \delta_{\lambda,\lambda'}  
\sum_{m}\sum_{\alpha,\alpha'}
(\mbox{\boldmath{$\alpha$}}_{\rm i})_{\alpha} 
 w(n\sigma;j_{\rm c}m;\alpha) \nonumber\\
&\times& w^*(n'\sigma';j_{\rm c}m;\alpha')
(\mbox{\boldmath{$\alpha$}}_{\rm f})_{\alpha'}, 
\label{eq.Mdef}
\end{eqnarray}
with $\xi=(\lambda,n,\sigma)$ and $\xi'=(\lambda',n',\sigma')$.
The summations for $\alpha$ and $\alpha'$ run over $x, y,$ and $z$.
The $\hat{M}(\mbox{\boldmath{$\alpha$}}_{\rm i},
             \mbox{\boldmath{$\alpha$}}_{\rm f};j_{\rm c})$ 
is regarded as a vector with 576 dimensions. 
The $\hat{Y}^{+-}(q)$ represents
the generalized density-density correlation function defined by
\begin{equation}
\left[\hat{Y}^{+-}({\bf q},\omega)\right]_{\xi_1\xi'_{1};\xi\xi'}
=  
\int_{-\infty}^{\infty} 
\langle [\rho_{{\bf q}\xi_1\xi'_1}(t)]^{\dagger}
       \rho_{{\bf q}\xi\xi'}(0)\rangle 
{\rm e}^{i\omega t}{\rm d}t,
\end{equation}
with
\begin{equation}
  \rho_{{\bf q}\xi\xi'} 
= (4/N)^{\frac{1}{2}}
\sum_{\bf k} d_{\xi}^{\dagger}([{\bf k+q}])d_{\xi'}({\bf k})
 {\rm e}^{-i{\bf G}\cdot{\bf a}_{\lambda}}.
\label{eq.density}
\end{equation}
Thereby $\hat{Y}^{+-}(q)$ is a matrix of $576\times 576$ dimensions. 
Since an extra phase factor 
${\rm e}^{-i{\bf G}\cdot{\bf a}_{\lambda}}$ is contained  in 
(\ref{eq.density}), the RIXS intensities may become different 
between the inside and the outside of the first MBZ.

\subsection{$\hat{Y}^{+-}(q)$ within the RPA}
Now that the RIXS spectra are connected with the density-density correlation
function, we outline the procedure to calculate it within the RPA.
See \cite{Igarashi2016} for details.

Let us introduce the two-particle Green's function,
\begin{equation}
\left[\hat{Y}^{{\rm T}}(q)\right]_{\xi_1\xi'_{1};\xi\xi'}
=-i\int \left\langle 
T\left\{[\rho_{{\bf q}\xi_1\xi'_{1}}(t)]^{\dagger}
\rho_{{\bf q}\xi\xi'}(0)\right\}\right\rangle
{\rm e}^{i\omega t}{\rm d}t,
\label{eq.green_time}
\end{equation}
where $T$ denoting the time ordering operator 
and $X(t)\equiv {\rm e}^{i Ht}X{\rm e}^{-i Ht}$.
By taking account of the multiple scattering between 
particle-hole pair, it is expressed as 
\begin{equation}
\hat{Y}^{{\rm T}}(q)
= \hat{F}(q)[\hat{I}+\hat{\Gamma}\hat{F}(q)]^{-1}
= \left[\hat{F}(q)^{-1}+\hat{\Gamma}\right]^{-1},
\label{eq.time_ladder}
\end{equation}
where $\hat{\Gamma}$ stands for the 
antisymmetrized vertex function, which is expressed as 
\begin{equation}
[\hat{\Gamma}]_{\xi_{2}\xi'_{2};\xi_{1}\xi'_{1}}=
g(\xi_{2}\xi'_{1};\xi_{1}\xi'_{2})-g(\xi_{2}\xi'_{1};\xi'_{2}\xi_{1}),
\end{equation}
with $g$ defined by the coefficient in the Coulomb interaction,
\begin{equation}
 H_{\rm I} = \frac{1}{2}\sum_{i}\sum_{\nu_1,\nu_2,\nu_3,\nu_4}
 g(\nu_1\nu_2;\nu_3\nu_4) d_{i\nu_1}^{\dagger}d_{i\nu_2}^{\dagger}
 d_{i\nu_4}d_{i\nu_3}.
\end{equation}
Then, $\hat{F}(q)$ in (\ref{eq.time_ladder}) is defined as
\begin{eqnarray}
&&\hspace*{-0.5cm} 
[\hat{F}(q)]_{\xi_2\xi'_{2};\xi_1\xi'_{1}} \equiv
-i \frac{4}{N}\sum_{{\bf k}}\int\frac{{\rm d}k_{0}}{2\pi}
[\hat{G}([{\bf k+q}],k_{0}+\omega)]_{\xi_{2}\xi_{1}}
\nonumber \\
&& \times [\hat{G}({\bf k},k_{0})]_{\xi'_{1}\xi'_{2}}
 {\rm e}^{i{\bf G}\cdot({\bf a}_{\lambda_2}-{\bf a}_{\lambda_1})},
\label{eq.free}
\end{eqnarray}
where $\hat{G}({\bf k},\omega)$ stands for the single-particle Green's 
function:
\begin{equation}
\left[\hat{G}({\bf k},\omega)\right]_{\xi,\xi'}
=-i \int\langle 
 T[d_{\xi}({\bf k},t)d_{\xi'}^{\dagger}({\bf k},0)]\rangle
 {\rm e}^{i \omega t}{\rm d}t.
\label{eq.dG}
\end{equation}
This may be written within the HFA as
\begin{equation}
[\hat{G}({\bf k},\omega)]_{\xi,\xi '}
=\sum_{\ell}
\frac{[\hat{U}({\bf k})]_{\xi,\ell}[\hat{U}({\bf k})^{-1}]_{\ell,\xi '}}
{\omega-E_{\ell}({\bf k})+i\delta{\rm sgn}[E_{\ell}({\bf k})]}, 
\label{eq.Green}
\end{equation}
where ${\rm sgn}[A]$ stands for a sign of quantity $A$ and 
$\delta$ denotes a positive convergent factor.
The $E_{\ell}({\bf k})$ and $[\hat{U}({\bf k})]_{\xi,\ell}$
stand for the $\ell$-th energy eigenvalue  measured from the chemical 
potential and the corresponding wave function, respectively.
By inserting (\ref{eq.Green}) into (\ref{eq.free}), we have
\begin{eqnarray}
&&\hspace*{-0.50cm} [\hat{F}(q)]_{\xi_2\xi'_{2};\xi_1\xi'_{1}} 
=  \frac{4}{N}\sum_{{\bf k}}\sum_{\ell,\ell '}
 U_{\xi_{2}\ell}({\bf k+q})U_{\xi_{1}\ell}^{*}({\bf k+q})
\nonumber\\
&\times& U_{\xi'_{1}\ell '}({\bf k})U_{\xi'_{2}\ell '}^{*}({\bf k})
 {\rm e}^{i{\bf G}\cdot({\bf a}_{\lambda_2}-{\bf a}_{\lambda_1})}
\nonumber\\
&\times& \left[\frac{[1-n_{\ell}({\bf k+q})]n_{\ell '}({\bf k})}
 {\omega-E_{\ell}({\bf k+q})+E_{\ell '}({\bf k})+i\delta}
\right. \nonumber\\
&& \left.
 -\frac{n_{\ell}({\bf k+q})[1-n_{\ell '}({\bf k})]}
 {\omega-E_{\ell}({\bf k+q})+E_{\ell '}({\bf k})-i\delta}\right].
 \label{eq.green_pair}
\end{eqnarray}
Once we obtain the two-particle Green's 
function $\hat{Y}^{{\rm T}}(q)$,
we can evaluate the density-density correlation function
$\hat{Y}^{+-}(q)$ 
with the help of the fluctuation-dissipation theorem,
\begin{equation}
 \left[\hat{Y}^{+-}(q)\right]_{\xi_1\xi'_{1};\xi\xi'}=
 i \left\{
 \left[\hat{Y}^{{\rm T}}(q)\right]_{\xi_1\xi'_{1};\xi\xi'}
-\left[\hat{Y}^{{\rm T}}(q)
 \right]^{*}_{\xi\xi';\xi_1\xi'_{1}}
\right\}.
\label{eq.fdt1}
\end{equation}

\section{Calculated results\label{sect.4}}
In addition to the continuous states of individual electron-hole
pair excitations, there emerge several bound states as poles in 
$\hat{Y}^{+-}(q)$. To obtain such bound states, we search for 
$\omega$ giving zero eigenvalue in $\hat{F}(q)^{-1}+\hat{\Gamma}$.
In this procedure, we evaluate $\hat{F}(q)$ by summing over 
${\bf k}$ in (\ref{eq.green_pair}) with dividing the MBZ into 
$40\times 40$ meshes. 
Let $\omega_{B}({\bf q})$ be the bound-state energy. Then
the residue of the pole, which is necessary to
calculate the spectral intensity, is evaluated by finite difference 
between $\omega=\omega_B({\bf q})$ and 
$\omega=\omega_B({\bf q})+0.0001$eV in place of the differentiation.
In order to compare the calculated RIXS spectra with those of
the experiment, we should specify the polarization setting.
As given in (\ref{eq.pol.i}) and (\ref{eq.pol.f}), 
the incident photon with $\pi$-polarization
is prepared, then, the scattered photons have been
collected without separating 
the $\sigma'$ and $\pi'$ polarizations
in the experiment\cite{Gretarsson2013-2}.
Thus, all the results shown in this section represent 
the sum of the spectra in the $\pi-\sigma'$ and $\pi-\pi'$ channels.

\subsection{Magnetic excitations}
\begin{figure}
\includegraphics[width=8.0cm]{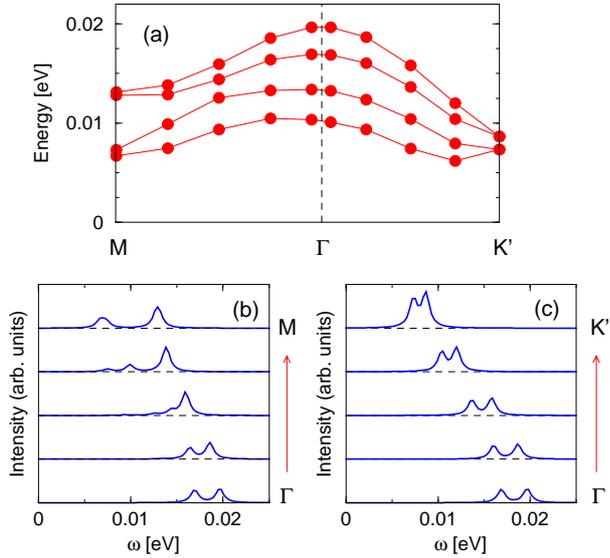}
\caption{\label{fig.int}
Panel (a): Dispersion relation of magnetic excitations for ${\bf q}$ 
along symmetry directions. Panels (b) and (c):
RIXS spectra in a horizontal scattering geometry
as a function of $\omega$ for ${\bf q}$ from the $\Gamma$ to the M points, 
and from the $\Gamma$ to the K' points.
Intensities of the $\sigma$ and $\pi$ polarizations are summed for the
scattered x-ray. The spectra are convoluted with the Lorentzian
function with half width of half maximum 1 meV.
}
\end{figure}

Panel (a) in figure \ref{fig.int} shows the calculated bound-state 
energy $\omega_{B}({\bf q})$ as a function of ${\bf q}$ along 
the symmetry directions in the low-energy region. 
We find that four gapped excitation modes reside below 20 
meV, which may be identified as magnetic excitations.
Such small energy scale is consistent with the energy difference between
the zigzag and other magnetic orders discussed in Sec. \ref{sect.2.2}.
The presence of four modes may be consistent with the spin-wave excitations
in the localized $j_{\rm eff}=\frac{1}{2}$ spin model in the zigzag ordered state.
At the K' point, four modes become two 
pairs of degenerate modes.
Experimentally, the magnetic excitations have been identified
below 6 meV by INS
\cite{Choi2012} and found extending up to 35 meV by RIXS 
\cite{Gretarsson2013-2}. 
It suggests there exist at least two magnetic excitation modes,
which is consistent with our results.
Finally, the excitation energy becomes smaller with ${\bf q}$ 
away from the $\Gamma$ point in qualitative agreement with 
the RIXS experiment\cite{Gretarsson2013-2}.

Panels (b) and (c) in figure \ref{fig.int} show the RIXS spectra 
as a function of
$\omega$ for ${\bf q}$ from the $\Gamma$ to the M points and the $\Gamma$
to K' points, respectively. The $\delta$-function peaks are convoluted with 
the Lorentzian function with the half width of half maximum 1 meV.
Among four modes, two upper-energy modes have intensities 
much larger than two lower-energy modes in the present scattering geometry.

\subsection{Excitonic excitations} 
Following the same procedure as the magnetic excitations,
we find sixteen bound states for $0.4<\omega_{B}({\bf q})<0.8$ eV 
below the energy continuum of individual electron-hole pair 
excitations in the density-density 
correlation function, which may be identified as excitonic excitations.
They are composed mainly of a pair of
$j_{\rm eff}=\frac{1}{2}$ electron and $j_{\rm eff}=\frac{3}{2}$ hole.
The presence of sixteen modes may be consistent with 
the localized excitations
from $j_{\rm eff}=\frac{3}{2}$ to $\frac{1}{2}$ states for four
Ir ions in the unit cell of the zigzag ordering state.
The continuous spectra are roughly estimated by
sorting $E_{\ell}({\bf k+q})-E_{\ell'}({\bf k})$ into segments 
with the width 0.05 eV in (\ref{eq.green_pair}), 
resulting in a histogram representation.

\begin{figure}
\includegraphics[width=8.0cm]{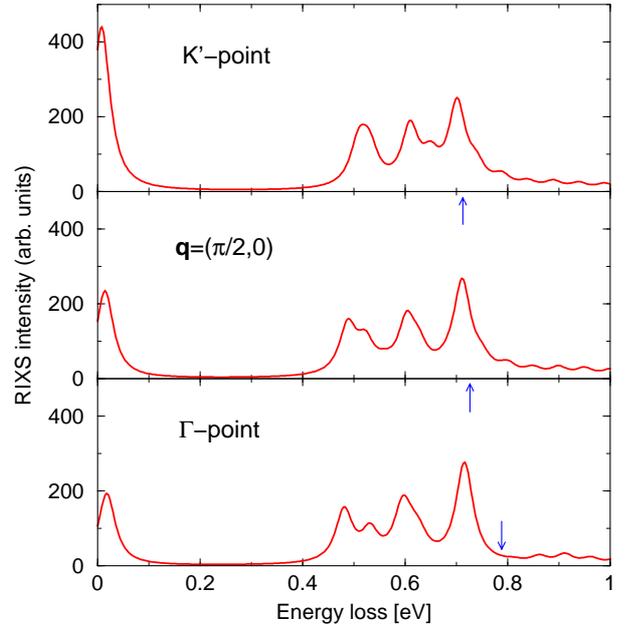}%
\caption{\label{fig.combined}
The RIXS spectra for both the magnon and the exciton excitations 
as a function of $\omega$ for ${\bf q}$ from the $\Gamma$ to $K'$ points.
Arrows indicate the lowest boundary of individual electron-hole pair 
excitations.
The spectra are convoluted with Lorentzian function
with the half width half maximum 0.04 eV. 
} 
\end{figure}

Figure \ref{fig.combined} shows the calculated RIXS spectra as a function 
of $\omega$ for ${\bf q}$ from the $\Gamma$ to the K' points.
The spectra are convoluted with the Lorentzian function with 
the half width of half maximum 0.04 eV. 
Arrows indicates the lowest boundary of the
energy continuum. We have three prominent peaks. 
In the RIXS experiment, three peaks named A, B, and C have been 
observed at 
$\omega\sim 0.42$ eV, $0.72$ eV, and $0.83$ eV, respectively, 
with little momentum dependence \cite{Gretarsson2013-1}.
The structure found in our calculated spectra seems to agree
qualitatively with that derived from the calculation on a small cluster 
\cite{BH.Kim2014} and the experimental spectra\cite{Gretarsson2013-1}.

\section{\label{sect.5}Concluding remarks}
We have analyzed RIXS spectra in Na$_2$IrO$_3$ on a multi-orbital 
tight-binding model composed only of the $t_{2g}$ orbitals for Ir ions 
on a honeycomb lattice. Using conventional parameter values for the SOI,
the Coulomb interaction, and the Ir-Ir transfer energy, we have carried
out the HFA to the model to calculate the one-electron energy as well as 
the ground-state energy by fixing the staggered magnetic moment 
along the $a$ axis. 
The zigzag ordering phase is found stabler than the N\'{e}el 
and the stripy ordering phases.
It may be remarkable that the HFA to the simple tight-binding model
leads to the zigzag order, since the energy difference from other orders
are rather small.

We have formulated the RIXS spectra in terms of the density-density 
correlation function within the FCA. 
The RIXS spectra have been evaluated within the RPA.
In the correlation function, there appear four bound states below 
20 meV, and sixteen bound states between 
0.4 and 0.8 eV, below the energy continuum of
the individual electron-hole pair excitations.
These bound states constitute the spectral peaks, which are in qualitative 
agreement with the RIXS experiment \cite{Gretarsson2013-1,Gretarsson2013-2}.
Note that the HFA-RPA scheme has worked well for describing
the excitation spectra in the presence of the antiferromagnetic 
long-range 
order in La$_2$CuO$_4$ \cite{Nomura2005,Nomura2015} and 
Sr$_2$IrO$_4$ \cite{Igarashi2014-3}. 
The present results are of qualitative nature, based on a simple 
model and neglecting electron correlations.
To be quantitative along the itinerant-electron approach, 
it may be necessary to use more realistic models, and to go beyond the HFA-RPA.

\begin{acknowledgments}
We thank M. Takahashi for valuable discussions.
This work was partially supported by a Grant-in-Aid for Scientific Research 
from the Ministry of Education, Culture, Sports, Science and Technology
of the Japanese Government.
\end{acknowledgments}

\bibliographystyle{apsrev}
\bibliography{paper2}

\end{document}